\documentclass[prd,aps,twocolumn,showpacs,preprintnumbers,amsmath,amssymb,nofootinbib,showkeys]{revtex4-1}

\RequirePackage[T1]{fontenc}

\usepackage{epsfig,graphicx,amssymb,amsmath}
\RequirePackage{mathptmx}      
\RequirePackage{flushend}
\RequirePackage{color}
\usepackage{braket}
\usepackage{lipsum}
\usepackage{changes}
\usepackage{footnote}
\usepackage{multirow}
\RequirePackage{hyperref}
\hypersetup{
	linktocpage,
    colorlinks,
    citecolor=blue,
    filecolor=black,
    linkcolor=blue,
    urlcolor=blue,
}

\begin{document}

\title{Assessment of the $a_{2}(1320)$ tensor-meson contribution to $\eta/\eta^{\prime}\to\pi^{0}\gamma\gamma$ decays}

\author{Rafel Escribano$^{1,2}$}
\email{rescriba@ifae.es}
\author{Sergi Gonz\`{a}lez-Sol\'{i}s$^{3,4}$}
\email{sergig@icc.ub.edu}
\author{Emilio Royo$^{2,5}$}
\email{emilio.royocarratala@uchceu.es}

\affiliation{
$^1$Grup de F\'{i}sica Te\`{o}rica, Departament de F\'{i}sica, 
Universitat Aut\`{o}noma de Barcelona, 08193 Bellaterra (Barcelona), Spain\\
$^2$Institut de F\'{i}sica d'Altes Energies (IFAE) and 
The Barcelona Institute of Science and Technology (BIST), 
Campus UAB, 08193 Bellaterra (Barcelona), Spain\\
$^3$Departament de F\'isica Qu\`{a}ntica i Astrof\'isica (FQA), 
Universitat de Barcelona (UB), c.~Mart\'i i Franqu\`{e}s, 1, 08028 Barcelona, Spain\\
$^4$Institut de Ci\`{e}ncies del Cosmos (ICCUB), 
Universitat de Barcelona (UB), c.~Mart\'i i Franqu\`{e}s, 1, 08028 Barcelona, Spain\\
$^5$Departamento de Matem\'{a}ticas, F\'{i}sica y Ciencias Tecnol\'{o}gicas, 
Universidad Cardenal Herrera-CEU, CEU Universities, 46115 Alfara del Patriarca, Val\`{e}ncia, Spain
}




\begin{abstract}
In light of the recent measurement of the $\eta\to\pi^{0}\gamma\gamma$ decay by the KLOE-2 Collaboration, 
a previous analysis including vector- and scalar-meson exchange contributions
using the VMD and L$\sigma$M frameworks, respectively, 
is extended in the present study to incorporate the effects of the $a_{2}(1320)$ tensor meson
within a chiral context.
Although the individual contribution of the $a_{2}$ is negligible,
its destructive interference with the vector-meson resonances is found to be significant,
representing approximately $18\%$ of the total signal and substantially affecting the 
diphoton invariant-mass distribution, especially at low $m_{\gamma\gamma}^{2}$ values.
The total decay rate is calculated to be $\Gamma(\eta\to\pi^{0}\gamma\gamma)=0.154(22)$~eV, 
which corresponds to a branching ratio of ${\rm BR}(\eta\to\pi^{0}\gamma\gamma)=1.17(17)\times10^{-4}$.
This result is approximately $5\sigma$ below the reported value of the PDG, $2.55(22)\times10^{-4}$,
while it is in very good agreement with the KLOE-2 measurement, $0.98(11_{\rm stat})(14_{\rm syst})\times10^{-4}$.
In contrast, the total contribution of the $a_{2}$ is found to be negligible in 
$\eta^{\prime}\to\pi^{0}\gamma\gamma$, 
as this process is completely dominated by the exchange of an on-shell $\omega$ vector resonance.
\end{abstract}



\maketitle

\section{Introduction}
\label{SectionIntroduction}
The doubly radiative process $\eta\to\pi^{0}\gamma\gamma$ is a very rare Standard Model meson decay channel
that, despite having a history of experimental measurements and theoretical predictions spanning several
decades~\cite{ParticleDataGroup:2024cfk,Achasov:2001qm,Gan:2020aco}, 
still remains a subject of controversy regarding the underlying decay mechanism and value of its decay width.

The experimental situation is far from conclusive, 
as significant discrepancies among existing measurements still exist.
The first notable measurement was reported by the GAMS-2000 Collaboration in 1984, 
with a branching ratio of $7.1(1.4)\times10^{-4}$~\cite{Serpukhov-Brussels-AnnecyLAPP:1984udf}
and an uncertainty of $20\%$. 
More recent experimental analyses by the A2 and Crystal Ball Collaborations
yield substantially lower values with improved accuracy.
The most precise measurement comes from the MAMI A2 experiment, 
which in 2014 and based on 1200 events reported 
${\rm{BR}}(\eta\to\pi^{0}\gamma\gamma)=2.52(23)\times10^{-4}$~\cite{A2atMAMI:2014zdf} 
with a $9\%$ error.
This value is consistent within uncertainties with the measurement
$2.21(24_{\rm stat})(47_{\rm syst})\times 10^{-4}$~\cite{Prakhov:2008zz}
obtained by Cristal Ball in 2008 with fewer statistics.
However, the very recent measurement by the KLOE-2 Collaboration based on 1246(133) events, 
${\rm{BR}}(\eta\to\pi^{0}\gamma\gamma)=
0.98(11_{\rm stat})(14_{\rm syst})\times10^{-4}$~\cite{KLOE-2:2025ggc},\footnote{This 
new measurement corroborates the older KLOE result, ${\rm{BR}}(\eta\to\pi^{0}\gamma\gamma)=
0.84(30)\times10^{-4}$~\cite{KLOE:2005hln}, based on a data sample of only 68(23) events.}
challenges the higher values reported by A2 and Crystal Ball, 
falling a factor of $2.6$ below the A2 measurement.

From a theoretical point of view, this decay presents a unique challenge to low-energy effective theories, 
as low-order contributions from chiral perturbation theory either vanish or are strongly suppressed, 
shifting the leading contributions to unusually high orders in the chiral
expansion~\cite{Ametller:1991dp,Ko:1993rg}. 
These higher-order effects are theoretically understood in terms of meson-resonance exchanges, 
with vector mesons playing a dominant role.
The implementation of these vector-meson exchanges has a significant impact on the
prediction of the decay width:
the result obtained with the full vector-meson propagator is a factor of two larger than the one
found by replacing the propagators by contact terms~\cite{Ametller:1991dp}. 
Tree-level contributions at $\mathcal{O}(p^{6})$ due to the scalar $a_{0}(980)$ and tensor $a_{2}(1320)$
resonances were also evaluated in Ref.~\cite{Ametller:1991dp},
but these had several sources of uncertainty, 
the most important being the signs of the $a_{0}$ and $a_{2}$ contributions,
which could not be unambiguously fixed.
The sign ambiguity of the scalar contribution was fixed by the chiral unitary approach of
Refs.~\cite{Oset:2002sh,Oset:2008hp}, in which the $a_{0}$ was not introduced into the amplitude
in an \textit{ad hoc} manner,
but rather dynamically, being generated through the coupled-channel rescattering of the
$\pi^{0}\eta$ and $K\bar{K}$ intermediate channels.
In Ref.~\cite{Escribano:2018cwg}, an all-order estimate of the scalar contributions was calculated
using the linear sigma model framework, which allowed assessing the relevance of the full scalar propagators. 
It should be highlighted that Refs.~\cite{Oset:2002sh,Oset:2008hp,Escribano:2018cwg} 
did not consider the contribution of the $a_{2}(1320)$ tensor meson. 
More sophisticated unitarization methods for generating the $a_{0}(980)$ 
are based on dispersive approaches applied to the crossing-symmetric amplitude of the production reaction
$\gamma\gamma\to\pi^{0}\eta$~\cite{Danilkin:2017lyn,Lu:2020qeo}. 
These methods have demonstrated good agreement with the $\gamma\gamma\to\pi^{0}\eta$ 
experimental data from the Belle Collaboration~\cite{Belle:2009xpa},
where the $a_{2}$, modeled using a Breit-Wigner parameterization, is also included.
The theoretical predictions in all these works exhibit a wide range of results, 
from ${\rm{BR}}(\eta\to\pi^{0}\gamma\gamma)=2.52(61)\times10^{-4}$, 
obtained using unitarized chiral perturbation theory~\cite{Oset:2008hp}, 
to ${\rm{BR}}(\eta\to\pi^{0}\gamma\gamma)=2.22(19)\times10^{-4}$~\cite{Danilkin:2017lyn} 
and ${\rm{BR}}(\eta\to\pi^{0}\gamma\gamma)=1.81^{+0.46}_{-0.33}\times10^{-4}$~\cite{Lu:2020qeo}, 
using dispersive methods, and ${\rm{BR}}(\eta\to\pi^{0}\gamma\gamma)=1.35(8)\times10^{-4}$, 
from a VMD and L$\sigma$M analysis~\cite{Escribano:2018cwg}.

More important than the partial width for understanding the underlying dynamics 
is the diphoton invariant-mass distribution.
It has long been put forward that the shape of the distribution of this decay is rather sensitive
to scalar dynamics and, in particular, to the interplay between vector- and scalar-resonance exchanges.
Ref.~\cite{Oset:2008hp} claimed that scalar contributions could lead to a significant enhancement
of the full distribution at high diphoton invariant masses,
but the present experimental uncertainties per bin in the spectrum prevents the determination
of the role played by scalars. 
These effects may be tested in the near future at the
Jefferson Lab Eta Factory Experiment (JEF)~\cite{JEF}, 
where a new distribution measurement is expected with enough precision to test
theoretical mechanisms and potentially distinguish the contribution of the vector-scalar interference
from the pure vector one.
In any case, the role of scalar resonances remains unclear~\cite{Gan:2020aco}, 
as the findings of Refs.~\cite{Lu:2020qeo,Danilkin:2017lyn,Escribano:2018cwg} 
have revised earlier expectations regarding their relevance, 
suggesting that their impact on future high-precision $\eta\to\pi^0\gamma\gamma$ measurements
may be less significant than previously thought:
Ref.~\cite{Lu:2020qeo} incorporates scalar dynamics in their analysis and finds
the corresponding contributions to be suppressed; 
Ref.~\cite{Danilkin:2017lyn} does not explicitly retain the scalar contributions in the decay region
and finds that they are not required to describe the measured spectrum by A2; 
Ref.~\cite{Escribano:2018cwg} also includes scalars in their analysis, but finds
their effect to be minimal,
with the spectrum instead dominated by VMD and showing good agreement with the recent KLOE-2 measurement. 

In this work, we evaluate the impact of the $a_{2}(1320)$ tensor-meson contribution 
to the decay $\eta\to\pi^{0}\gamma\gamma$.
Our goal is to perform a dedicated study of this effect and show that future high-precision measurements
are likely to be more sensitive to the vector-tensor interference than to the vector-scalar one.
For the present analysis, we extend our previous VMD and L$\sigma$M amplitudes of 
Ref.~\cite{Escribano:2018cwg} by incorporating the $a_{2}(1320)$ tensor-meson exchange
to the decay amplitude.
We find that the $a_{2}(1320)$ contribution decreases the decay width by approximately $14\%$ 
compared to our previous result \cite{Escribano:2018cwg}. 
We also investigate the contribution of the $a_{2}(1320)$ to the partner decay
$\eta^{\prime}\to\pi^{0}\gamma\gamma$. 
In this case, the exchange of an on-shell $\omega$ resonance is by far the dominant mechanism, 
resulting in a negligible contribution of the $a_{2}$.

This article is structured as follows.
In Sec.~\ref{SectionVMD}, 
we briefly review the theoretical frameworks for the vector- and scalar-meson exchange contributions,
and, in Sec.~\ref{Sectiona2}, we introduce the formalism to include the
$a_{2}(1320)$ tensor-meson contribution.
In Sec.~\ref{SectionResults}, we present our predictions for the decay widths 
and diphoton invariant-mass distributions, as well as a discussion of the results.
We conclude this work with a summary and conclusions in Sec.~\ref{SectionSummary}.\\

\section{Theoretical framework}\label{SectionFramework}
\subsection{Vector- and scalar-meson exchange contributions}\label{SectionVMD}

The vector- and scalar-meson exchange contributions can be calculated within the frameworks of vector meson dominance
(VMD) and the linear sigma model (L$\sigma$M), respectively.
We briefly review these models here and refer the reader to Ref.~\cite{Escribano:2018cwg} for further details.

The vector-meson contributions to the $\eta\to\pi^{0}\gamma\gamma$ proceed through the decay sequence
$\eta\to V\gamma$ followed by $V\to\pi^{0}\gamma$, 
with a total of six diagrams contributing to the amplitude, 
corresponding to the exchange of the vector-meson resonances $V=\rho^{0},\omega$ and $\phi$
in the $t$- and $u$-channels.
Combining the vertices $V\eta\gamma$ and $V\pi^{0}\gamma$
with the propagator of the exchanged vector resonances, 
one arrives at the VMD amplitude~\cite{Escribano:2018cwg}
\begin{widetext}
\begin{equation}
\label{AVMDetapi0}
\quad {\cal A}^{\mathrm{VMD}}_{\eta\to\pi^0\gamma\gamma}=
\sum_{V=\rho^0, \omega, \phi}g_{V\!\eta\gamma}g_{V\!\pi^0\gamma}
\left[\frac{(P\cdot q_2-m_\eta^2)\{a\}-\{b\}}{D_V(t)}
+\left\{
\begin{array}{c}
q_2\leftrightarrow q_1\\
t\leftrightarrow u
\end{array}
\right\}\right]\ ,
\end{equation}
\end{widetext}
where
$t,u=(P-q_{2,1})^2=m_\eta^2-2P\cdot q_{2,1}$ are Mandelstam variables,
$\{a\}$ and $\{b\}$ are the Lorentz structures 
\begin{equation}
\label{LorentzStruct}
\begin{aligned}
\{a\}&=(\epsilon_1\cdot\epsilon_2)(q_1\cdot q_2)-(\epsilon_1\cdot q_2)(\epsilon_2\cdot q_1)\ ,\\[1ex]
\{b\}&=(\epsilon_1\cdot q_2)(\epsilon_2\cdot P)(P\cdot q_1)+(\epsilon_2\cdot q_1)
(\epsilon_1\cdot P)(P\cdot q_2)\\
&-(\epsilon_1\cdot\epsilon_2)(P\cdot q_1)(P\cdot q_2)-(\epsilon_1\cdot P)(\epsilon_2\cdot P)
(q_1\cdot q_2)\ ,
\end{aligned}
\end{equation}
with $P$ being the four-momentum vector of the $\eta$, and $\epsilon_{1,2}$ 
and $q_{1,2}$ the polarization and four-momentum vectors of the photons, respectively. 

The denominator $D_V(q^2)$ is the propagator of the exchanged vector meson.
For the $\rho^0$ meson, we use $D_{\rho^{0}}(q^2)=m_{\rho^0}^2-q^2-i\,m_{\rho^0}\Gamma_{\rho^0}(q^{2})$,
with the energy-dependent decay width given by
\begin{equation}
\label{varGamma}
\Gamma_{\rho^0}(q^2)=\Gamma_{\rho^0}
\left(\frac{q^2-4m_{\pi^{\pm}}^2}{m_{\rho^0}^2-4m_{\pi^{\pm}}^2}\right)^{3/2}\theta(q^2-4m_{\pi^{\pm}}^2)\ ,
\end{equation}
whereas for the $\omega$ and $\phi$ mesons the usual Breit-Wigner propagator with constant width is used. 
The amplitude of the decay $\eta^\prime\to\pi^0\gamma\gamma$
is analogous to that of $\eta\to\pi^0\gamma\gamma$ in Eq.~(\ref{AVMDetapi0}), 
with the replacements $m_\eta^2\to m_{\eta^\prime}^2$ 
and $g_{V\eta\gamma}g_{V\pi^0\gamma}\to g_{V\eta^\prime\gamma}g_{V\pi^0\gamma}$.

We note that while the most general $VP\gamma$ couplings in Eq.~(\ref{AVMDetapi0}) are energy dependent, 
\textit{i.e.}~$g_{VP\gamma}(q^2)$,
in practice this energy dependence vanishes as the photons are on-shell
and the corresponding couplings become just constants~\cite{Escribano:2020rfs,Escribano:2022njt}.
For the current analysis, we determine the values of the $g_{VP\gamma}$ 
couplings directly from experiment rather than relying on any specific 
model~\cite{Bramon:1994pq,Prades:1993ys,Escribano:2018cwg,Escribano:2020jdy}, 
rendering our treatment of the vector exchanges model independent. 
Using the most up-to-date experimental data from the PDG~\cite{ParticleDataGroup:2024cfk}, 
we find the values for the $g_{VP\gamma}$ couplings summarized in
Table~\ref{gVPgammacouplingsempirical}.\footnote{It should be noted that, from the theoretical expressions for the decay widths and the experimental data, 
one can only determine the absolute value of the $g_{VP\gamma}$ couplings, 
which is what we show in Table~\ref{gVPgammacouplingsempirical}. 
However, one can determine the sign of the couplings by making use of phenomenological models, 
such as the one discussed in Refs.~\cite{Bramon:1997va,Escribano:2020jdy}. 
All $g_{VP\gamma}$ couplings are found to be positive,
except for $g_{\phi\eta^\prime\gamma}$, which turns out to be negative.}
\begin{table}[h]
\centering
\begin{tabular}{|lll|}
\hline
Decay &\qquad BR &\qquad $|g_{V\!P\gamma}|$ GeV$^{-1}$\\
\hline
$\rho^0\to\pi^{0}\gamma$		&\qquad $(4.7\pm0.8)\times10^{-4}$    &\qquad $0.22(2)$\\ 
$\rho^0\to\eta\gamma$			&\qquad $(3.00\pm0.21)\times10^{-4}$  &\qquad $0.48(2)$\\ 
$\eta^{\prime}\to\rho^0\gamma$	&\qquad $(29.48\pm0.35)\%$              &\qquad $0.393(7)$\\ 
$\omega\to\pi^{0}\gamma$		&\qquad $(8.33\pm0.25)\%$             &\qquad $0.71(1)$\\ 
$\omega\to\eta\gamma$			&\qquad $(4.5\pm0.4)\times10^{-4}$    &\qquad $0.136(6)$\\ 
$\eta^{\prime}\to\omega\gamma$	&\qquad $(2.52\pm0.07)\%$             &\qquad $0.122(2)$\\ 
$\phi\to\pi^{0}\gamma$			&\qquad $(1.33\pm0.05)\times10^{-3}$  &\qquad $0.041(1)$\\ 
$\phi\to\eta\gamma$				&\qquad $(1.306\pm0.024)\%$           &\qquad $0.2096(20)$\\ 
$\phi\to\eta^{\prime}\gamma$	&\qquad $(6.23\pm0.21)\times10^{-5}$  &\qquad $0.216(4)$\\ 
\hline     
\end{tabular}
\caption{PDG values~\cite{ParticleDataGroup:2024cfk} for the branching ratios of the 
$V(P)\to P(V)\gamma$ transitions and the corresponding $g_{V\!P\gamma}$ couplings.} 
\label{gVPgammacouplingsempirical}
\end{table}

Concerning the scalar-meson exchange contributions to $\eta/\eta^{\prime}\to\pi^{0}\gamma\gamma$, 
we refer the reader to Ref.~\cite{Escribano:2018cwg} for details. 
In this reference, the scalar contributions are evaluated using the L$\sigma$M 
and a complete one-loop propagator is employed to describe the propagation of the scalar resonances. 
This analysis shows that scalar contributions are subdominant, 
while vector-resonance exchanges largely saturate the signal.

\subsection{Contribution of the $a_{2}(1320)$}\label{Sectiona2}

\begin{figure}[ht!]
\centering\includegraphics[scale=0.65]{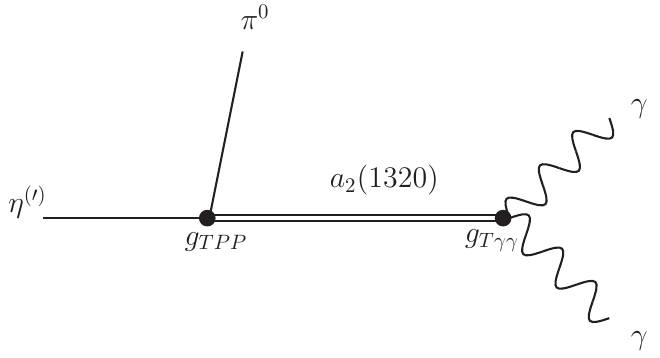}
\caption{Schematic diagram for the $a_{2}$-exchange contribution to the decay 
$\eta^{(\prime)}\to\pi^{0}\gamma\gamma$.}
\label{Fig:a2Exchange} 
\end{figure}

The Feynman diagram depicting the intermediate $a_{2}$-exchange contribution to the decay $\eta\to\pi^{0}\gamma\gamma$ is shown in Fig.~\ref{Fig:a2Exchange}.
In order to couple tensor mesons to pseudo-Nambu-Goldstone bosons (pNGB), 
we follow the formalism presented in Ref.~\cite{Ecker:2007us}.

The chiral Lagrangian required to describe the decay of a tensor meson into two pseudoscalars is given by
\begin{equation}
\label{Eq:LagTPP}
\mathcal{L}_{TPP}=g_{TPP}\langle T_{\mu\nu}\left\{u^{\mu},u^{\nu}\right\}\rangle\ ,
\end{equation}
where $T_{\mu\nu}$ is the matrix of tensor-meson fields, 
which reads
\begin{equation}
T_{\mu\nu}=\begin{pmatrix}\frac{a_{2}^{0}}{\sqrt{2}}+\frac{f_{2}^{8}}{\sqrt{6}}
+\frac{f_{2}^{0}}{\sqrt{3}}&a_{2}^{+}&K_{2}^{*+}\cr
a_{2}^{-}&-\frac{a_{2}^{0}}{\sqrt{2}}+\frac{f_{2}^{8}}{\sqrt{6}}+\frac{f_{2}^{0}}{\sqrt{3}}&K_{2}^{*0}\cr
K_{2}^{*-}&\bar{K}_{2}^{*0}&-\frac{2f_{2}^{8}}{\sqrt{6}}+\frac{f_{2}^{0}}{\sqrt{3}}\end{pmatrix}_{\mu\nu}\ ,
\end{equation}
and $u_\mu$ is defined as
\begin{equation}
u_{\mu}=i\left(u^{\dagger}\partial_{\mu}u-u\partial_{\mu}u^{\dagger}\right)\ , 
\end{equation} 
with the usual choice of coset representative $u=e^{i\frac{\Phi}{\sqrt{2}F_{\pi}}}$
and pNGB field matrix 
\begin{equation}
\Phi=
\begin{pmatrix}
\frac{\pi^{0}}{\sqrt{2}}+\frac{\eta_{8}}{\sqrt{6}}+\frac{\eta_{0}}{\sqrt{3}}&\pi^{+}&K^{+}\cr
\pi^{-}&-\frac{\pi^{0}}{\sqrt{2}}+\frac{\eta_{8}}{\sqrt{6}}+\frac{\eta_{0}}{\sqrt{3}}&K^{0}\cr
K^{-}&\bar{K}^{0}&-\frac{2\eta_{8}}{\sqrt{6}}+\frac{\eta_{0}}{\sqrt{3}}
\end{pmatrix}\ .
\end{equation}

The partial decay width $T\to P_{1}P_{2}$ can be calculated from Eq.~(\ref{Eq:LagTPP}) 
making use of the polarization sum for tensors~\cite{Ecker:2007us}
\begin{equation}
\sum_{\rm{pol}}\epsilon_{\mu\nu}(q)\epsilon^{*}_{\rho\sigma}(q)=P_{\mu\nu,\rho\sigma}(q)\ ,
\end{equation}
where
\begin{equation}
\begin{aligned}
P_{\mu\nu,\rho\sigma}(q)&=\frac{1}{2}\left(P_{\mu\rho}P_{\nu\sigma}
+P_{\nu\rho}P_{\mu\sigma}\right)-\frac{1}{3}P_{\mu\nu}P_{\rho\sigma}\ ,\cr
P_{\mu\nu}&=g_{\mu\nu}-\frac{q_{\mu}q_{\nu}}{m_{a_{2}}^{2}}\ .
\end{aligned}
\end{equation}
Consequently, we find the following decay widths for the two processes $a_{2}\to\pi\eta^{(\prime)}$
\begin{equation}
\label{Eq:Decaya2pieta}
\Gamma(a_{2}\to\pi\eta)=\frac{8 p^{5}(m_{a_{2}},m_{\pi},m_{\eta})}{15\pi m_{a_{2}}^{2}}
\frac{|g_{TPP}|^{2}}{F_{\pi}^{4}}\cos^2\phi\ ,
\end{equation}
and
\begin{equation}
\label{Eq:Decaya2pietaprime}
\Gamma(a_{2}\to\pi\eta^\prime)=\frac{8 p^{5}(m_{a_{2}},m_{\pi},m_{\eta^\prime})}{15\pi m_{a_{2}}^{2}}
\frac{|g_{TPP}|^{2}}{F_{\pi}^{4}}\sin^2\phi\ ,
\end{equation}
with $\phi$ being the $\eta$-$\eta^{\prime}$ mixing angle in the quark-flavor basis~\cite{Bramon:1997va}, 
and the three-momentum $p(m_{T},m_{1},m_{2})$ in the decaying-particle rest frame is given by
\begin{equation}
\begin{aligned}
&p(m_{T},m_{1},m_{2})=\frac{m_{T}}{2}\cr
&\ \times\sqrt{\left[1-(m_{1}+m_{2})^{2}/m_{T}^{2}\right]
\left[1-(m_{1}-m_{2})^{2}/m_{T}^{2}\right]}\ .
\end{aligned}
\end{equation}
Using Eqs.~(\ref{Eq:Decaya2pieta}) and (\ref{Eq:Decaya2pietaprime}), 
together with the total decay width $\Gamma_{a_{2}}=107(5)$ MeV
and the branching ratios ${\rm{BR}}(a_{2}\to\pi\eta)=14.5(1.2)\%$ 
and ${\rm{BR}}(a_{2}\to\pi\eta^{\prime})=5.5(9)\times10^{-3}$~\cite{ParticleDataGroup:2024cfk}, 
we find, respectively, the coupling strengths
\begin{equation}
|g_{TPP}|=21.5(1.0)\,{\rm{MeV}}\ ,
\label{Eq:CouplingEtaPiEta}
\end{equation}
and
\begin{equation}
|g_{TPP}|=22.4(1.9)\,{\rm{MeV}}\ .
\label{Eq:CouplingEtaPiEtaP}
\end{equation}
Next, we investigate the decay of the $a_{2}$ into two photons using the
chiral Lagrangian~\cite{Bellucci:1994eb,Chen:2023ybr} 
\begin{equation}
\label{Eq:LagTgammagamma}
\mathcal{L}_{T\gamma\gamma}=c_{T\gamma\gamma}\langle T_{\mu\nu}\Theta_{\gamma}^{\mu\nu}\rangle\ ,
\end{equation}
where\footnote{There is some inconsistency in the literature regarding the definition of $\Theta_{\gamma}^{\mu\nu}$.
In Refs.~\cite{Bellucci:1994eb,Chen:2023ybr}, 
it is written as in Eq.~\eqref{Eq:LagTgammagammaTheta} of this work,
while Ref.~\cite{Giacosa:2005bw} employs 
\begin{equation}
\nonumber\Theta_{\gamma}^{\mu\nu}= f_{+\alpha}^{\quad\mu}f_{+}^{\alpha\nu}+\frac{1}{4}g^{\mu\nu}f_{+}^{\rho\sigma}f_{+\rho\sigma}\ ,
\end{equation}
which differs in the ordering of the indices in the first factor, 
leading to a relative sign due to the antisymmetry of the electromagnetic tensor.
Moreover, some authors (see, e.g., Refs.~\cite{Drechsel:1999rf,Danilkin:2017lyn}) 
retain only the first term in $\Theta_{\gamma}^{\mu\nu}$, while Ref.~\cite{Lu:2020qeo} employs a different interaction $\mathcal{L}_{T\gamma\gamma}$.
We have carefully examined the symmetric energy–momentum tensor and find that the correct form is that given in Ref.~\cite{Bellucci:1994eb}.
}
\begin{equation}
\label{Eq:LagTgammagammaTheta}
\Theta_{\gamma}^{\mu\nu}=
f_{+\alpha}^\mu f_{+}^{\alpha\nu}+\frac{1}{4}g^{\mu\nu}f_{+}^{\rho\sigma}f_{+\rho\sigma}\ ,
\end{equation}
and
\begin{equation}
\begin{aligned}
f_{+}^{\mu\nu}&=e\left(uQu^{\dagger}+u^{\dagger}Qu\right)F^{\mu\nu}\ ,\cr
F^{\mu\nu}&=\partial^{\mu}A^{\nu}-\partial^{\nu}A^{\mu}\ ,
\end{aligned}
\end{equation}
with $A^\mu$ being the electromagnetic field, $F^{\mu\nu}$ the electromagnetic field strength tensor, 
and $Q={\rm{diag}}\{2/3,-1/3,-1/3\}$ the quark-charge matrix. 
From Eq.~(\ref{Eq:LagTgammagamma}), 
we obtain the expression for the $a_{2}\to\gamma\gamma$ partial width\footnote{Note that the second term in Eq.~(\ref{Eq:LagTgammagammaTheta}) 
does not contribute for an on-shell tensor.}
\begin{equation}
\Gamma(a_{2}\to\gamma\gamma)=\frac{8\pi}{45}\alpha^{2}m_{a_{2}}^{3}|c_{T\gamma\gamma}|^{2}\ ,
\end{equation}
which allows us to determine 
\begin{equation}
|c_{T\gamma\gamma}|=1.2\times 10^{-4}\ {\rm{MeV}}^{-1}\ ,
\label{Eq:CouplingTensorGammaGamma}
\end{equation}
with a negligible error, 
by making use of the corresponding experimental branching ratio, 
${\rm{BR}}(a_{2}\to\gamma\gamma)=9.4(7)\times10^{-6}$~\cite{ParticleDataGroup:2024cfk}.

Now, we can combine the $a_{2}\pi^{0}\eta^{(\prime)}$ and $a_{2}\gamma\gamma$ vertices
from Eqs.~(\ref{Eq:LagTPP}) and (\ref{Eq:LagTgammagamma}),
respectively, with the tensor propagator for the $a_2$~\cite{Kubis:2015sga} (cf.~Fig.~\ref{Fig:a2Exchange})
\begin{equation}
\begin{split}
i\,\frac{P_{\mu\nu,\rho\sigma}(s)}
       {m_{a_2}^2 - s - i\, m_{a_2}\Gamma_{a_2}(s)}\ ,
\end{split}
\label{EqPropagatora2}
\end{equation}
\smallskip

\noindent
where $\Gamma_{a_{2}}(s)$ is the energy-dependent width of the $a_2$ resonance,
which in turn is given by~\cite{Belle:2009xpa}
\begin{equation}
\label{Eq:EnergyDependentWidtha2}
\Gamma_{a_2}(s)=\sum_X \Gamma_{X_1X_2(X_3)}(s)\ ,
\end{equation}
with $X_1X_2(X_3)$ standing for $\rho\pi$, $\eta\pi$, $K\bar{K}$, $\gamma\gamma$, and $\omega\pi\pi$. 
The partial width $\Gamma_{X_1X_2}$ is parameterized as \cite{Belle:2009xpa,Blatt:1952ije}
\begin{equation}
\begin{aligned}
\label{Eq:BlattWeisskopf}
\Gamma_{X_1X_2}(s)=&\,
\Gamma_{a_2}{\rm{BR}}(a_2\to X_1X_2)\cr
&\times\left[\frac{q_X(s)}{q_X(m_{a_2}^2)}\right]^5\frac{D_2(q_X(s)r_{a_2})}{D_2(q_X(m_{a_2}^2)r_{a_2})}\ ,
\end{aligned}
\end{equation}
where $\Gamma_{a_2}$ is the total decay width at the resonance mass,
$q_X(s)=\sqrt{s}/2\times\sqrt{[1-(m_{X_1}+m_{X_2})^2/s][1-(m_{X_1}-m_{X_2})^2/s]}$, the
Blatt-Weisskopf barrier factors are defined as $D_2(x)=1/(x^4+3x^2+9)$,
and $r_{a_2}$ is an effective interaction radius which is fitted in Ref.~\cite{Belle:2009xpa}
and found to be $3.09^{+0.53}_{-0.55}$\ (GeV/c)$^{-1}$. 
For the three-body decay, $\Gamma_{\omega\pi\pi}(s)=\Gamma_{a_2}{\rm{BR}}(a_2\to\omega\pi\pi) s/m_R^2$
is used in place of Eq.~(\ref{Eq:BlattWeisskopf}).

The invariant amplitude for the tensor contribution to the $\eta\to\pi^{0}\gamma\gamma$ decay can, thus, be written as
\begin{widetext}
\begin{equation}
\begin{aligned}
\label{Tensoretapi0}
\quad{\cal A}^{a_{2}}_{\eta\to\pi^0\gamma\gamma}=&
\frac{32e^{2}}{3F_{\pi}^{2}}g_{TPP}\,c_{T\gamma\gamma}\cos\phi\,
P^{\mu}p^{\nu}\frac{P_{\mu\nu,\rho\sigma}(s)}{m_{a_{2}}^{2}-s-im_{a_{2}}\Gamma_{a_{2}}(s)}\cr
&\times\Big\{g^{\rho\alpha}q_{1}^{\beta}q_{2}^{\sigma}+g^{\sigma\beta}q_{1}^{\rho}q_{2}^{\alpha}
-g^{\rho\alpha}g^{\sigma\beta}(q_{1}\cdot q_{2})-g^{\alpha\beta}q_{1}^{\rho}q_{2}^{\sigma}
+\frac{1}{2}g^{\rho\sigma}\big[g^{\alpha\beta}(q_{1}\cdot q_{2})-q_{1}^{\beta}q_{2}^{\alpha}\big]\Big\}
\epsilon_{\alpha}^{*}(q_{1})\epsilon_{\beta}^{*}(q_{2})\ ,
\end{aligned}
\end{equation}
\end{widetext}
where $P(p)$ is the four-momentum of the $\eta(\pi^0)$. 
This amplitude is fully determined, up to the overall sign of the product of coupling constants
$g_{TPP}\,c_{T\gamma\gamma}$.\footnote{This 
point is discussed further in the following section.}
Similarly, the amplitude for the contribution of the $a_{2}$ to the decay
$\eta^{\prime}\to\pi^{0}\gamma\gamma$ can be obtained from Eq.~(\ref{Tensoretapi0}) 
with the replacements $m_\eta\to m_{\eta^\prime}$ and $\cos\phi\to\sin\phi$.

\section{Results and discussion}\label{SectionResults}

\begin{table}[t!]
\begin{tabular}{|ll|}\hline
Term & Value (in eV)\\
\hline
Vectors (V) & 0.1661\\
Scalars (S) & 0.0005\\
$a_{2}$ & 0.0020\\
V-S interference & 0.0121\\
V-$a_{2}$ interference & $-0.0270$\\
$\Gamma_{\rm{th}}$ &0.1538
\\
\hline
BR$_{\rm{th}}$  &   $1.17(17)\times 10^{-4}$\\
BR$_{\rm{exp}}$ (PDG~\cite{ParticleDataGroup:2024cfk})  &   $2.55(22)\times 10^{-4}$\\
BR$_{\rm{exp}}$ (KLOE-2~\cite{KLOE-2:2025ggc})  &   $0.98(11_{\rm stat})(14_{\rm syst})\times 10^{-4}$\\
\hline
\end{tabular}
\caption{Individual contributions and total prediction for the $\eta\to\pi^{0}\gamma\gamma$ decay width 
(upper part). 
Comparison of our branching ratio with experimental measurements (lower part).}
\label{Table:DecayWidthPredictions}
\end{table}

With the amplitude derived in the previous section, we are now ready to present our results. 
These are obtained using the standard formula for three-body decays~\cite{ParticleDataGroup:2024cfk},
with the total amplitude written as the coherent sum of the vector-, scalar- and tensor-exchange contributions, 
$\mathcal{A}=\mathcal{A}_{\rm{VMD}}+\mathcal{A}_{\rm{L\sigma M}}+\mathcal{A}_{a_{2}}$. 

First, a remark on the sign of the product of couplings $g_{TPP}\,c_{T\gamma\gamma}$
of the tensor-exchange amplitude is in order.
While this sign is irrelevant for the standalone contribution of the $a_{2}(1320)$,
it becomes crucial when assessing its interference with the other exchanged resonances. 
To fix it, we make use of the crossing-symmetric amplitude for the scattering process
$\gamma\gamma\to\pi^{0}\eta$ evaluated in the relevant kinematic region.
We find that only the positive sign provides an unambiguous and satisfactory description of the $a_{2}$ line shape in the $\gamma\gamma\to\pi^{0}\eta$ scattering data from the Belle Collaboration~\cite{Belle:2009xpa}.
This conclusion is further supported when our amplitude is fitted to the experimental data, since the positive sign yields a clearly improved fit.\footnote{Specifically, we performed a fit to the Belle $\gamma\gamma\to\pi^{0}\eta$ 
scattering data, leaving the relative phase between 
the tensor contribution, $\mathcal{A}_{a_{2}}$, and the vector-scalar part, $\mathcal{A}_{V+S}$, as a free parameter.
The best-fit value obtained was approximately $0^\circ$, 
corresponding to the positive sign discussed in the main text.}
We note that this choice translates into a destructive interference in the decay region. 
\begin{figure*}[ht!]
\centering\includegraphics[scale=0.75]{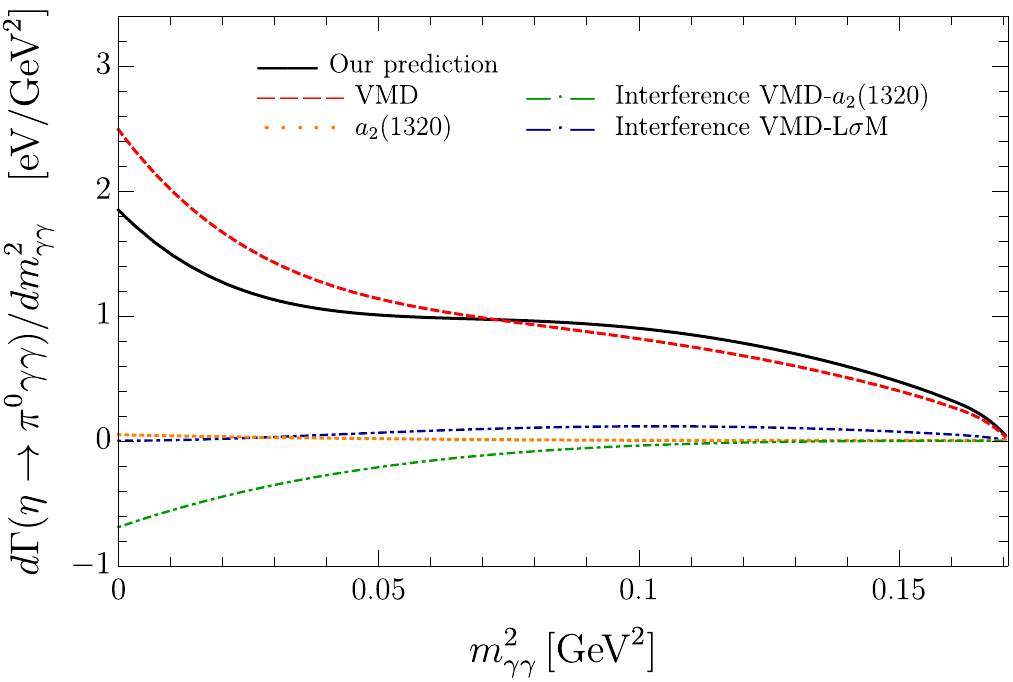}
\caption{Total (black solid) and individual contributions to the $\eta\to\pi^{0}\gamma\gamma$ 
invariant-mass distribution $d\Gamma/dm_{\gamma\gamma}^{2}$.}
\label{Fig:Predictions} 
\end{figure*}

\begin{figure*}[ht!]
\includegraphics[scale=0.55]{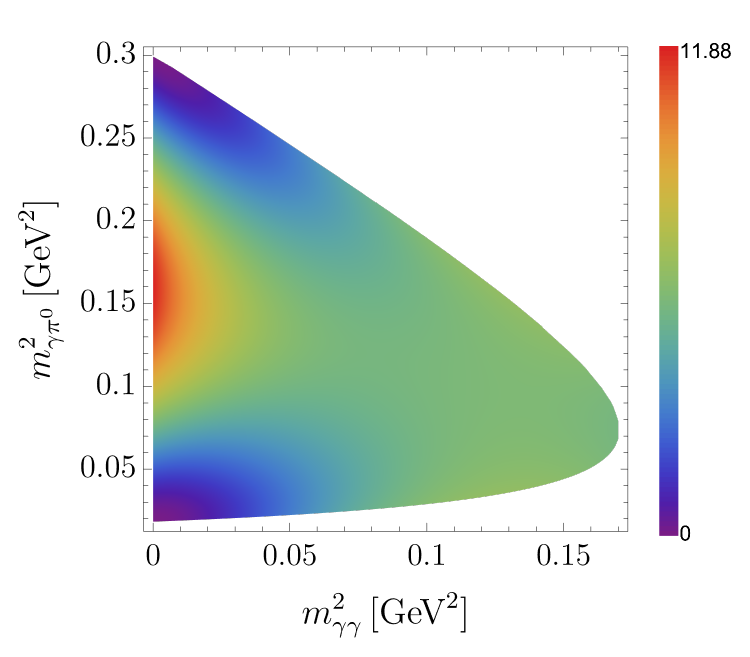}
\includegraphics[scale=0.55]{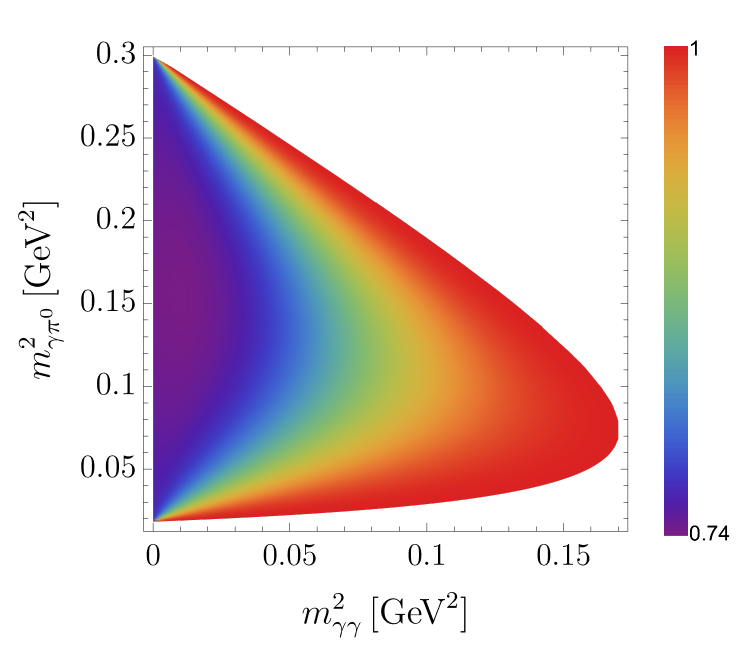}\\
\quad\includegraphics[scale=0.53]{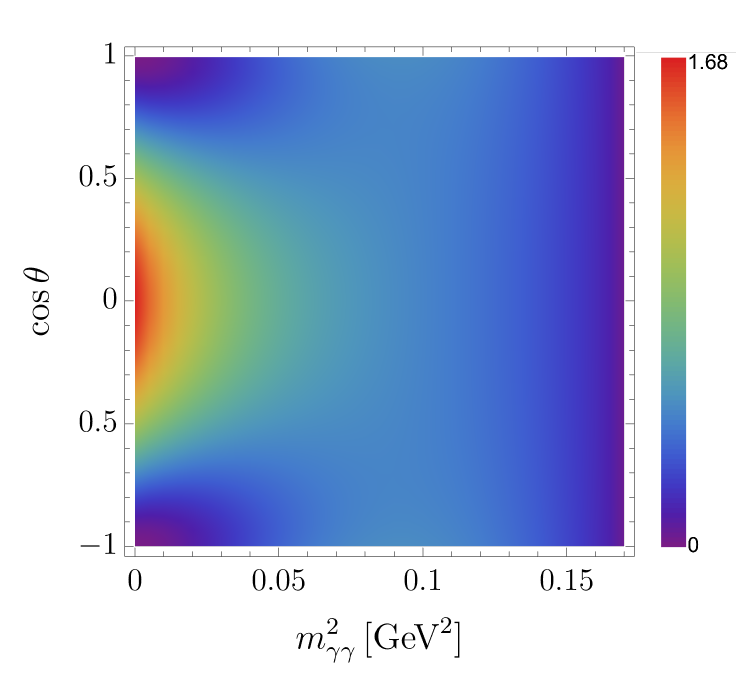}\quad
\includegraphics[scale=0.53]{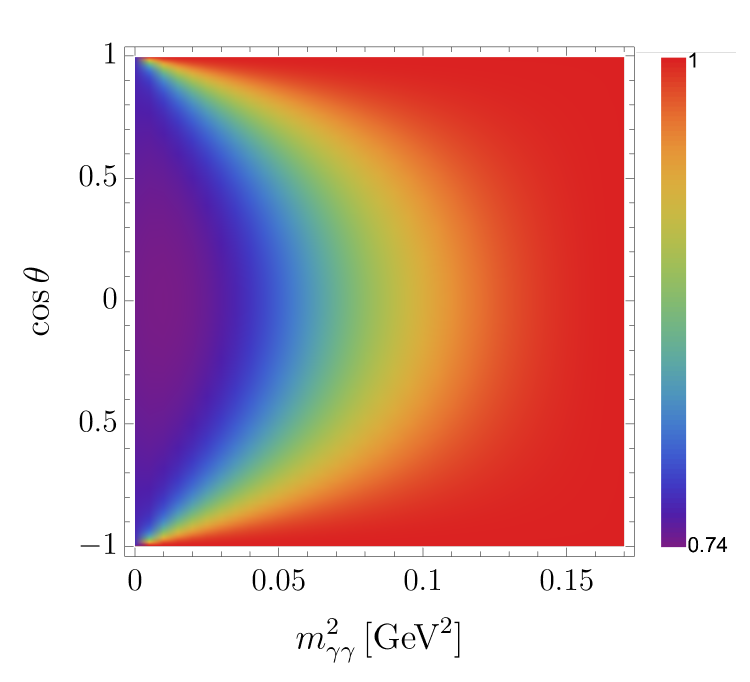}
\caption{Dalitz plots for the full squared amplitude $|\mathcal{A}_{{\rm{V+S+}}a_{2}}|^{2}$
of the decay $\eta\to\pi^{0}\gamma\gamma$ (left)
and for the ratio $|\mathcal{A}_{{\rm{V+S+}}a_{2}}|^{2}/|\mathcal{A}_{{\rm{V+S}}}|^{2}$ (right).}
\label{Fig:Dalitz} 
\end{figure*}

In Table~\ref{Table:DecayWidthPredictions},
we provide our numerical results for the individual contributions to the decay width coming from \mbox{vector-,}
scalar,- and tensor-meson exchanges, along with their corresponding interferences.
The total decay width is also presented, $\Gamma(\eta\to\pi^{0}\gamma\gamma)=0.154(22)$ eV, 
where the quoted error of approximately $14\%$ is due to the uncertainties of the VMD and tensor couplings.
As can be seen, the contribution from the exchange of vector mesons dominates with about $108\%$
of the total signal.
The second most important contribution arises from the destructive interference between the $a_{2}(1320)$ tensor meson
and the vector mesons, which represents approximately $18\%$ of the signal. This constitutes the main result of this work.
The remaining contributions come from the interference term between the vectors and the scalar,
accounting for $8\%$, from the individual $a_{2}(1320)$-exchange contribution, representing approximately $1.3\%$ of the signal, and from the scalar meson exchanges, which accounts for about $0.3\%$.\footnote{At
first glance, one might naively expect the contribution from the scalar exchange to exceed that 
of the tensor exchange, based on the masses of the corresponding resonances and the available phase space. 
However, it is important to note that regions of phase space with decreasing $m_{\gamma\gamma}^2$
are kinematically enhanced, with the collinear photon final-state configuration, $m_{\gamma\gamma}^2=0$, 
being strongly favored.
In conjunction with this, 
the Lorentz structure associated with a spin-0 exchange enforces a vanishing squared amplitude at
$m_{\gamma\gamma}^2=0$, which results in an additional suppression of the scalar contribution.
This effect is not observed in higher-spin exchanges because their Lorentz structures are richer.}
Finally, the scalar-tensor interference is insignificant and can be completely neglected.
In this table, we also show the integrated branching ratio along with the corresponding experimental values.
We predict $\rm{BR}(\eta\to\pi^{0}\gamma\gamma)=1.17(17)\times10^{-4}$,
representing a decrease of approximately $14\%$ with respect to the value obtained when excluding the contribution of the
$a_{2}(1320)$, $1.36(18)\times 10^{-4}$~\cite{Escribano:2018cwg}.\footnote{
The branching ratio from Ref.~\cite{Escribano:2018cwg} has been updated using the latest $VP\gamma$ couplings (cf. Table~\ref{gVPgammacouplingsempirical}), and the associated error has been corrected to account for the interference between the three vector mesons, which had previously been omitted.
}
A comparison of our prediction with experimental measurements shows that it is about
$5\sigma$ below the value reported by the PDG, $2.55(22)\times 10^{-4}$~\cite{ParticleDataGroup:2024cfk}, while it is only
approximately $0.8\sigma$ above, and hence in perfect agreement with, the recent measurement by the KLOE-2 Collaboration,
$0.98(11_{\rm stat})(14_{\rm syst})\times10^{-4}$~\cite{KLOE-2:2025ggc}.
Furthermore, compared with other recent theoretical calculations,
our prediction is about $1.5\sigma$ and $4.4\sigma$ away from the results from Ref.~\cite{Lu:2020qeo}, $\Gamma(\eta\to\pi^{0}\gamma\gamma)=0.237^{+0.060}_{-0.043}$ eV, and Ref.~\cite{Danilkin:2017lyn}, $\Gamma(\eta\to\pi^{0}\gamma\gamma)=0.291(22)$ eV, respectively.\footnote{Note that
Refs.~\cite{Lu:2020qeo,Danilkin:2017lyn} consider the decay $\eta\to\pi^{0}\gamma\gamma$
from dispersive amplitudes originally developed to describe the scattering $\gamma\gamma\to\pi^{0}\eta$ data 
from the Belle Collaboration~\cite{Belle:2009xpa}, with the $a_{2}$ resonance modeled as a Breit-Wigner.}
On the other hand, our central value shows a discrepancy of about $50\%$ with respect to the chiral unitarized result 
$\Gamma(\eta\to\pi^{0}\gamma\gamma)=0.33(8)$ eV~\cite{Oset:2008hp}
obtained neglecting the contribution of the $a_{2}$ resonance.
As noted, both the experimental and theoretical situations remain inconclusive,
making future measurements, such as the forthcoming measurement from the JEF experiment at JLab~\cite{JEF},
crucial for resolving the issue.\footnote{We will return to this point later.} 
\begin{figure*}[ht!]
\includegraphics[scale=0.5]{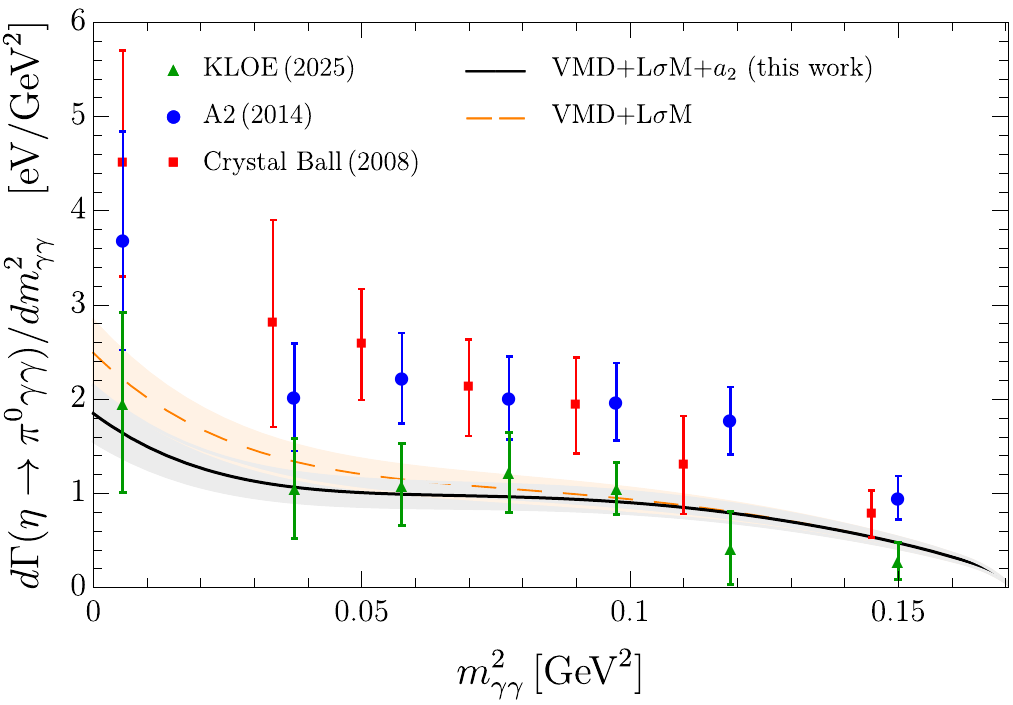}
\quad
\includegraphics[scale=0.5]{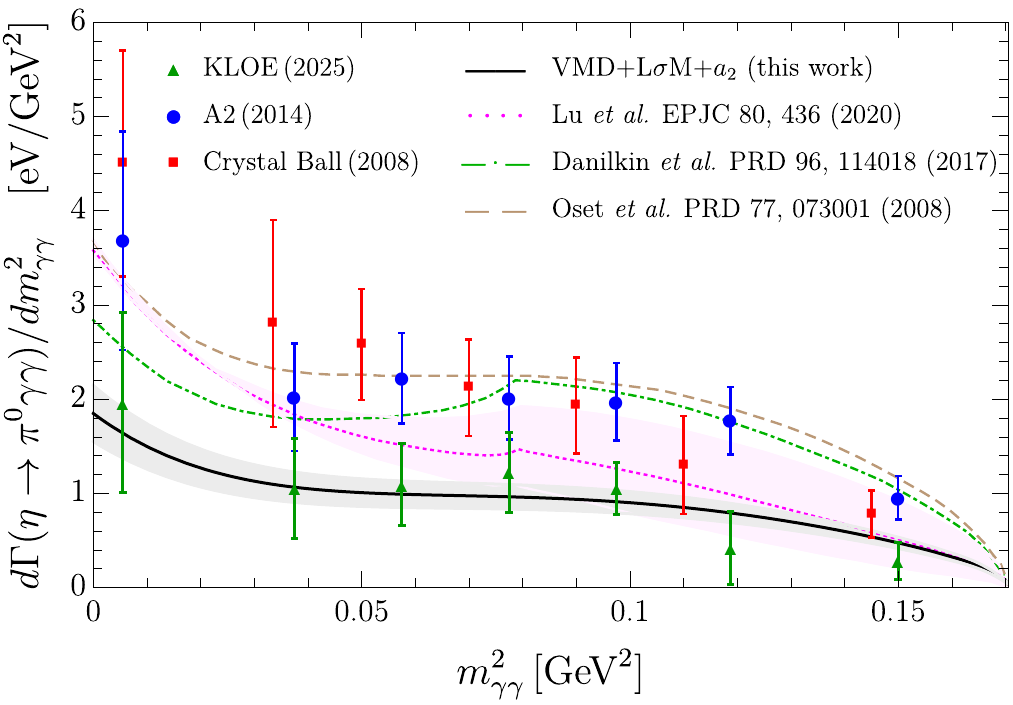}
\caption{{\it{Left}}:
Comparison between the A2 (blue circles)~\cite{A2atMAMI:2014zdf},
Crystal Ball (red squares)~\cite{Prakhov:2008zz}
and KLOE-2 (green triangles)~\cite{KLOE-2:2025ggc} measurements
of the $\eta\to\pi^{0}\gamma\gamma$ invariant-mass distribution, $d\Gamma/dm_{\gamma\gamma}^{2}$,
and our prediction (black solid line) based on vector-, scalar-, and tensor-meson exchanges.
The orange dashed line is the prediction from Ref.~\cite{Escribano:2018cwg},
where the contribution of the $a_{2}(1320)$ was not included.
{\it{Right}}:
Our prediction (black solid) compared to data, and to the theoretical results from 
Refs.~\cite{Danilkin:2017lyn} (magenta dotted), \cite{Lu:2020qeo} (green dash-dotted) and \cite{Oset:2008hp} (brown dashed).}
\label{Fig:PredictionsData}
\end{figure*}

We next analyze the shape of the decay spectrum. 
The different contributions to the $m_{\gamma\gamma}^{2}$ invariant-mass distribution are shown in
Fig.~\ref{Fig:Predictions}.\footnote{The
standalone scalar and the scalar-tensor interference contributions are so small 
(cf.~Table~\ref{Table:DecayWidthPredictions}) that they have been omitted in the figure.}
Vector-meson exchanges (dashed red line) clearly dominate the entire spectrum.
However, the contribution of the destructive vector-tensor interference (dot-dashed green line) remains significant.
In contrast, the comparatively small individual tensor contribution (dotted orange line)
is only noticeable at low diphoton invariant masses.
The doubly-differential decay widths shown in Fig.~\ref{Fig:Dalitz}
provide further information on the tensor contribution: in the upper- and lower-left panels we present, respectively, the Dalitz plots of the total squared amplitude  
$|\mathcal{A}_{{\rm{V+S+}}a_{2}}|^{2}$ in terms of $(m_{\gamma\gamma}^{2},m_{\gamma\pi^{0}}^{2})$
and $(m_{\gamma\gamma}^{2},\cos\theta)$,\footnote{$\theta$
is the angle between the $\pi^{0}$ and one photon in the diphoton rest frame.}
whereas in the right-hand side panels, we display the ratio $|\mathcal{A}_{{\rm{V+S+}}a_{2}}|^{2}/|\mathcal{A}_{{\rm{V+S}}}|^{2}$, plotted in terms of the same variables, highlighting
the effect of the tensor-meson resonance, which appears as a reduction of the distribution around 
$m_{\gamma\pi^{0}}^{2}\sim 0.15$ GeV$^{2}$ and $\cos\theta\sim0$.

\begin{figure*}[ht!]
\includegraphics[scale=0.5]{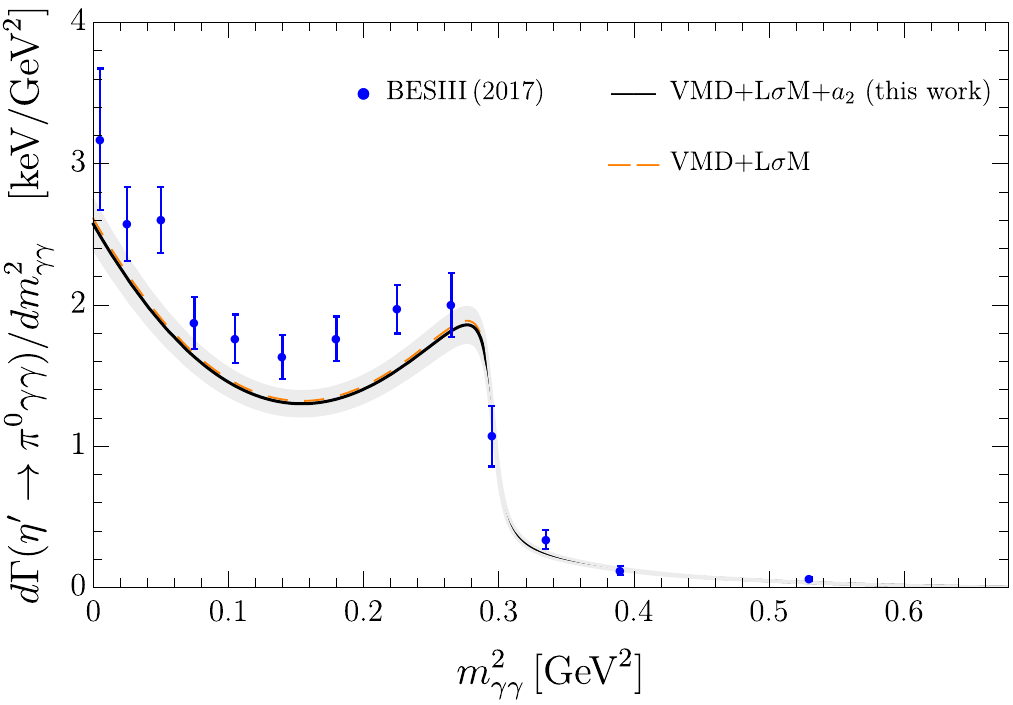}
\qquad
\includegraphics[scale=0.55]{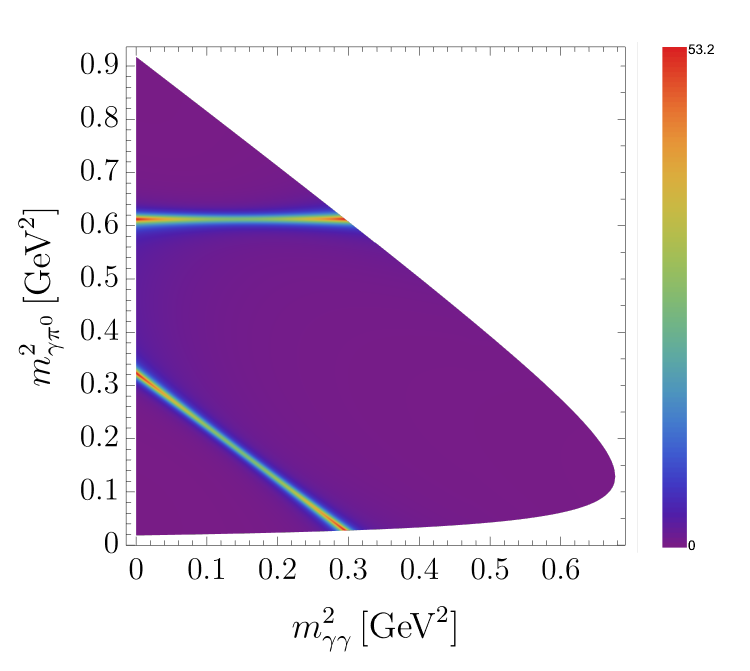}
\caption{{\it{Left}}:
Comparison between the BESIII (blue circles)~\cite{BESIII:2016oet} measurement
of the $\eta^{\prime}\to\pi^{0}\gamma\gamma$ invariant-mass distribution, $d\Gamma/dm_{\gamma\gamma}^{2}$,
and our prediction (black solid line) based on vector-, scalar-, and tensor-meson exchanges.
The orange dashed line is the prediction from Ref.~\cite{Escribano:2018cwg},
where the contribution of the $a_{2}(1320)$ was not included.
{\it{Right}}:
Dalitz plot for the decay $\eta^{\prime}\to\pi^{0}\gamma\gamma$.}
\label{Fig:PredictionsDataEtaP} 
\end{figure*}

In Fig.~\ref{Fig:PredictionsData},
we compare our predictions for the $m_{\gamma\gamma}^{2}$ invariant-mass distribution (black solid line)
with the experimental data from the A2~\cite{A2atMAMI:2014zdf}, Crystal Ball~\cite{Prakhov:2008zz}
and KLOE-2~\cite{KLOE-2:2025ggc} Collaborations (left plot),
and with the theoretical predictions from Refs.~\cite{Danilkin:2017lyn,Lu:2020qeo,Oset:2008hp} (right plot).
The left-hand side plot also shows the result from Ref.~\cite{Escribano:2018cwg},
updated with the most recent radiative couplings, where the $a_{2}$ contribution is not included (orange dashed line).
It is important to note the discrepancy of around $2\sigma$ at low $m_{\gamma\gamma}^{2}$ between
the black solid and orange dashed lines, which highlights the importance of the tensor-meson contribution.
It can be seen that our prediction is in excellent agreement with the recent KLOE-2 measurement, whereas
it is approximately a factor of 2 smaller than the data from A2 and Crystal Ball across the spectrum.
In addition, it deviates from the dispersive analysis of Ref.~\cite{Lu:2020qeo} (magenta dotted line in the right plot) at low diphoton masses, with this discrepancy reduced to within errors for large $m_{\gamma\gamma}^{2}$, while the deviation with respect to the result of Ref.~\cite{Danilkin:2017lyn} (green dash-dotted line) persists across the spectrum.\footnote{The cusp at $m_{\gamma\gamma}^2\approx 0.078$~GeV$^2$ in the analyses of Refs.~\cite{Danilkin:2017lyn,Lu:2020qeo}
corresponds to the $S$-wave $\pi^+\pi^-$ contribution to the process amplitude,
which is suppressed by isospin conservation.
At the amplitude level, we work in the isospin limit and, thus, this effect is absent in our computations.
By inspecting the spectrum of Lu \textit{et al.}~\cite{Lu:2020qeo},
one sees that this effect is small, and neglecting it is therefore justified.
An explanation of why this effect is relatively large in the result
of Danilkin \textit{et al.}~\cite{Danilkin:2017lyn} can be found in Ref.~\cite{Lu:2020qeo}.}
The prediction from Ref.~\cite{Oset:2008hp}, which does not include the $a_{2}$ contribution explicitly, is also shown for completeness (brown dashed line).
The discrepancies between experimental results, as well as theoretical calculations,
strongly motivate new measurements in ongoing experiments,
such as JEF~\cite{JEF}, or future $\eta$-factories, 
{\it{e.g.}},~the proposed REDTOP experiment~\cite{REDTOP:2022slw}, 
the super-$\tau$-charm facility~\cite{Achasov:2023gey}, or the $\eta$-factory at HIAF~\cite{Chen:2024wad}.
In fact, the projected measurements by the JEF experiment,
with an estimated uncertainty of approximately $5\%$ after 100 days of beam time~\cite{JEF,Gan:2020aco}, 
could be sufficient to assess the vector-tensor interference and distinguish it, for the first time,
from pure vector contributions.

Although the $a_{2}$ exchange contribution plays a significant role in $\eta\to\pi^{0}\gamma\gamma$
and can be probed with precise experimental measurements,
the decay $\eta^{\prime}\to\pi^{0}\gamma\gamma$
is overwhelmingly dominated by the VMD mechanism~\cite{Escribano:2018cwg},
in particular by the exchange of an $\omega$ resonance,
which can go on-shell due to the available phase space.
The VMD predictions from Ref.~\cite{Escribano:2018cwg} yield
$\rm{BR}(\eta^{\prime}\to\pi^{0}\gamma\gamma)=2.82(20)\times 10^{-3}$ and
$\rm{BR}(\eta^{\prime}\to\pi^{0}\gamma\gamma)=3.57(25)\times 10^{-3}$,
depending on whether the VMD couplings are determined directly from experiment,  with the most recent radiative couplings,
or fixed using a phenomenological model.
These values are compatible with the experimental result from BESIII,
$\rm{BR}(\eta^{\prime}\to\pi^{0}\gamma\gamma)=3.20(7)_{\rm stat}(23)_{\rm syst}
\times 10^{-3}$~\cite{BESIII:2016oet}, and
including the contribution of the $a_{2}(1320)$ resonance reduces the branching ratio by only about $1.4\%$,
yielding $\rm{BR}(\eta^{\prime}\to\pi^{0}\gamma\gamma)=2.78(21)\times 10^{-3}$
for the empirical-coupling case,
which is negligible compared to the experimental uncertainty of approximately $8\%$.
The diphoton spectrum predicted by VMD is displayed in the left-hand plot of
Fig.~\ref{Fig:PredictionsDataEtaP} (orange dashed line),
providing a satisfactory description of the distribution measured by BESIII~\cite{BESIII:2016oet}.
The black solid line includes, in addition, the contribution from the tensor exchange,
yielding a result that is only marginally smaller than the pure VMD prediction
and fully compatible within uncertainties.
This confirms that the effect of $a_{2}$ in this process is negligible at the current level of precision.
As a result, the associated Dalitz distribution (cf.~right-hand plot in Fig.~\ref{Fig:PredictionsDataEtaP})
does not provide information on the tensor resonance,
with the two narrow bands corresponding to the effect of the $\omega$ vector resonance.

\section{Summary and conclusions}\label{SectionSummary}

In this work,
we have performed a detailed analysis of the $a_{2}(1320)$-exchange contribution to the decay $\eta\to\pi^{0}\gamma\gamma$,
which is primarily driven by vector-meson exchanges,
with subleading scalar contributions~\cite{Escribano:2018cwg,Gan:2020aco}.
Our results show that, although the individual contribution of the $a_{2}(1320)$ is negligible,
its destructive interference with vector-meson resonances plays a significant role (cf.~Fig.~\ref{Fig:Predictions})
and noticeably modifies the diphoton invariant-mass distribution, 
particularly at low $m_{\gamma\gamma}^{2}$ values (cf.~Fig.~\ref{Fig:PredictionsData}).
This effect decreases the decay rate to $\Gamma(\eta\to\pi^{0}\gamma\gamma)=0.154(22)$ eV, corresponding to a branching ratio of $\rm{BR}(\eta\to\pi^{0}\gamma\gamma)=1.17(17)\times 10^{-4}$
(cf.~Table~\ref{Table:DecayWidthPredictions}), which is $14\%$ smaller than our previous result without the tensor contribution~\cite{Escribano:2018cwg}.
Our new prediction is in excellent agreement with the recent measurement from the KLOE-2 Collaboration~\cite{KLOE-2:2025ggc}.
\begin{figure}
\centering\includegraphics[scale=0.675]{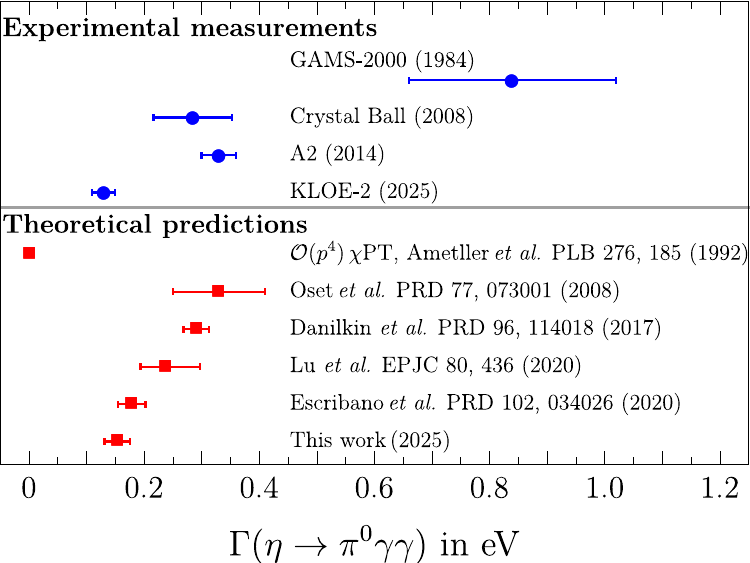}
\caption{Status of experimental measurements of the decay width $\Gamma(\eta\to\pi^{0}\gamma\gamma)$
(upper panel), including the results from the
GAMS-2000~\cite{Serpukhov-Brussels-AnnecyLAPP:1984udf}, 
Crystal Ball~\cite{Prakhov:2008zz},
A2~\cite{A2atMAMI:2014zdf},
and KLOE-2~\cite{KLOE-2:2025ggc}
Collaborations,
compared to theoretical predictions of $\chi$PT at $\mathcal{O}(p^{4})$~\cite{Ametller:1991dp},
Oset {\it{et al.}}~\cite{Oset:2008hp},
Danilkin {\it{et al.}}~\cite{Danilkin:2017lyn},
Lu {\it{et al.}}~\cite{Lu:2020qeo},
and Escribano {\it{et al.}}~\cite{Escribano:2018cwg} (using updated empirical $VP\gamma$ couplings).}
\label{Fig:DecayWidthExpvsTh} 
\end{figure}
Fig.~\ref{Fig:DecayWidthExpvsTh}
summarizes the state-of-the-art experimental measurements and theoretical calculations
for the partial decay width $\Gamma(\eta\to\pi^{0}\gamma\gamma)$.
The inconclusive situation provides strong motivation for new and more precise measurements,
such as those at the JEF experiment at JLab~\cite{JEF},
which may shed light on the interplay of meson resonances and,
in particular, distinguish the vector-tensor interference from the pure VMD mechanism.
In addition, improved measurements at future super $\eta/\eta^{\prime}$-factories,
such as REDTOP~\cite{REDTOP:2022slw} or HIAF~\cite{Chen:2024wad},
or at the proposed super tau-charm facility~\cite{Achasov:2023gey},
will provide deeper insight into the underlying dynamics.

We have also extended the analysis of the $\eta^{\prime}\to\pi^{0}\gamma\gamma$ decay by
including the $a_{2}$ exchange.
However, this process is overwhelmingly dominated by the exchange of vector resonances,
in particular the exchange of an intermediate virtual $\omega$ that can resonate due to
the available phase space.
The contribution of the $a_{2}$ results in an insignificant decrease in the branching ratio of approximately $1.4\%$. 

Finally, our findings motivate the study of tensor-meson contributions in semileptonic decays
$\eta/\eta^{\prime}\to\pi^{0}\ell^{+}\ell^{-}$ $(\ell=e,\mu)$~\cite{Escribano:2020rfs,Schafer:2023qtl},
which proceed through $\eta/\eta^{\prime}\to\pi^{0}\gamma^{*}\gamma^{*}$
followed by the conversion $\gamma^{*}\gamma^{*}\to\ell^{+}\ell^{-}$.
In addition, the inclusion of tensor-meson exchanges may help further constrain the parameter space of new hypothetical dark-sector particles, such as leptophobic vector bosons~\cite{Escribano:2022njt,Tulin:2014tya,Balytskyi:2021lzh}
or scalar particles~\cite{Gan:2020aco,Delaunay:2025lhl}, in the MeV-GeV mass range.
These investigations are left for future work.

\acknowledgments
The authors thank José Clavero for his participation during the preliminary stages of this work
and Alejandro Miranda for fruitful discussions.
The work of R.~E.~and E.~R.~is supported by 
the European Union’s Horizon 2020 Research and Innovation Programme 
under Grant No.~824093 (H2020-INFRAIA-2018-1) and by
the Ministerio de Ciencia, Innovación y Universidades under Grant No.~PID2023–146142NB-I00. 
The work of R.~E.~is also supported by the Departament de Recerca i Universitats 
from Generalitat de Catalunya to the Grup de Recerca ``Grup de Física Teòrica UAB/IFAE''
(Codi: 2021 SGR 00649).
R.~E.~also acknowledges financial support from the Spanish Ministry of Science and Innovation (MICINN) 
through the Spanish State Research Agency, under Severo Ochoa Centres of Excellence Programme 2025-2029 (CEX2024001442-S).
IFAE is partially funded by the CERCA program of the Generalitat de Catalunya. 
The work of E.~R.~is additionally funded by the Universidad Cardenal Herrera-CEU (INDI24/17 and GIR24/16).
The work of S.~G-S.~is supported by MICIU/AEI/10.13039/501100011033 and 
by FEDER UE through grant PID2023-147112NB-C21, as well as through the award
``Unit of Excellence María de Maeztu 2025-2029'' to the Institute of Cosmos Sciences, grant CEX2024-001451-M. 
Additional support is provided by the Generalitat de Catalunya (AGAUR) 
through grant 2021SGR01095. 
R.~E.~and S.~G-S.~are Serra H\'{u}nter Fellows.

\bibliography{refs}

\end{document}